\documentclass[runningheads,fleqn]{svmult}

\usepackage{makeidx}   
\usepackage{graphicx}  
\usepackage{subeqnar}  
\usepackage{multicol}  
\usepackage{physmult}  
\makeindex             

\begin{document}
\title*{Some Statistical Physics Approaches\\ for Trends and Predictions in
Meteorology}
\toctitle{Some Statistical Physics Approaches\protect\newline 
for Trends and Predictions 
in
Meteorology}
%
%
\titlerunning{Trends and Predictions in Meteorology}
\author{Kristinka Ivanova\inst{1} \and Marcel Ausloos\inst{2} 
\and Thomas Ackerman\inst{3} \and Hampton Shirer\inst{1} \and Eugene Clothiaux\inst{1}}

\authorrunning{K. Ivanova, et al}

\institute{Pennsylvania State University, University Park PA 16802, USA \and 
SUPRAS \& GRASP, B5, University of Li$\grave e$ge, B-4000 
Li$\grave e$ge, Belgium \and Pacific Northwest National Laboratory, Richland, 
WA 99352, USA }

\maketitle
\begin{abstract}

Specific aspects of time series analysis are discussed. They are related to
the analysis of atmospheric data that are pertinent to clouds. A brief
introduction
on some of the most interesting topics of current research on climate/weather
predictions is given.
Scaling properties of the liquid water path in stratus clouds are analyzed
to demonstrate the application of several methods of statistical physics for
analyzing data in atmospheric sciences, and more generally in geophysics.
The breaking up of a stratus cloud is shown to be related to changes in
the type of
correlations in the fluctuations of the signal that represents the total
vertical amount of liquid water in the stratus cloud. It is demonstrated that
the correlations of the liquid water path fluctuations exist indeed in a
more complex way than
usually known through their multi-affine dependence.
\end{abstract}

\section{Introduction}

Earth's climate is determined by complex interactions between sun, oceans,
atmosphere, land and biosphere \cite{andrews,atmosphere}. The 
composition of the
atmosphere is particularly important because certain gases, including water
vapor, carbon dioxide, etc., absorb heat radiated from the Earth's surface. As the
atmosphere warms, it in turn radiates heat back to the surface that 
increases the
earth's mean surface temperature by some 30~K above the value that 
would occur in
the absence of a radiation-trapping atmosphere \cite{andrews}. Perturbations in
the concentration of these radiatively active gases alter the intensity of this
effect on the earth's climate.

Climate change, a major concern of everyone, is a focus of current atmospheric
research. Understanding the processes and properties that effect atmospheric
radiation and, in particular, the influence of clouds and the role of cloud
radiative feedback, are issues of scientific interest. This leads to efforts to
improve not only models of the earth's climate but also predictions of climate
change \cite{clouds,beer}, whence weather prediction and climate models.

Lorenz's \cite{Lorenz} famous pioneering work on chaotic systems 
using a
simple set of nonlinear differential equations  was motivated by
considerations of weather prediction. However, predicting the results 
of  complex
nonlinear interactions that are taking place in an open system is a difficult
task. Yet physicists have only the Navier-Stokes equations \cite{LandauLifshitz} at
hand for describing fluid motion, in terms of such quantities as mass, pressure, temperature,
humidity, velocity, energy exchange, ... whence for describing
the variety of processes that take place in the atmosphere. Since controlled
experiments cannot be performed on the climate system, we rely on use of models
to identify cause-and-effect relationships. It is also essential to concentrate
on predicting the uncertainty in forecast models of weather and climate
\cite{Palmer,Philander}.

Modeling the impact of clouds is difficult because of their complex 
and differing
effects on weather and climate. Clouds can reflect incoming sunlight and,
therefore, contribute to cooling, but they also absorb infrared 
radiation leaving
the earth and contribute to warming. High cirrus clouds, for example, may have
the impact of warming the atmosphere. Low-lying stratus clouds, which are
frequently found over oceans, can contribute to cooling. In order to 
successfully
model and predict climate, we must be able to both describe the 
effect of clouds
in the current climate and predict the complex chain of events that 
might modify
the distribution and properties of clouds in an altered climate.

Much attention has been paid recently \cite{physworld} to the importance of the
main substance of the atmosphere and clouds, water in its three forms ---
vapor, liquid and solid, for buffering the global temperature against 
reduced or
increased solar heating \cite{ou}. Owing to its special properties, it is
believed, that water establishes lower and upper boundaries on how far the
temperature can drift from current values.

The role of clouds and water vapor in climate change is not well understood; yet
water vapor is the most abundant greenhouse gas and directly affects  
cloud cover and
the propagation of radiant energy. In fact, there may be positive feedback
between water vapor and other greenhouse gases. Carbon dioxide and other gases
from human activities slightly warm the atmosphere, increasing its ability to
hold water vapor. Increased water vapor can amplify the effect of an 
incremental
increase of other greenhouse gases.

Other studies suggest that the heliosphere influences the climate on Earth via
global mechanism that affects cloud cover \cite{marsh,svens}. Surprisingly the
influence of solar variability is found to be strongest in low clouds (3 km),
which points to a microphysical mechanism involving aerosol formation that is
enhanced by ionization due to cosmic rays.

Beyond the scientifically sound and highly sophisticated computer models, there
is still space for simple approaches, based on standard statistical physics
techniques and ideas, in particular based on the scaling hypothesis
\cite{schroeder}, phase transitions \cite{StanleyPTbook} and percolation theory
aspects \cite{StaufferP2book}. Analogies can be found between 
meteorological and
other phenomena in social or natural science \cite{Bakbook}. However to
distinguish cases and patterns due to ''external field'' influences or
self-organized criticality \cite{Turcotte} is not obvious indeed. The coupling
between human activities and deterministic physics is hard to model on simple
terms.

There have been several reports that long-range power-law correlations can be
extracted from apparently stochastic time series in meteorology
\cite{bunde,bunde2} and multi-affine properties 
\cite{turbwavelet,mfatmoturb} can
be identified related to atmospheric turbulence \cite{atmoturb}. The 
same type of
investigations has already appeared and seems promising in atmospheric science.
In the following we touch upon a brief review of some statistical physics
approaches for testing scaling hypothesis in meteorology and for identifying the
self-affine or multi-affine nature of atmospheric quantities. We apply useful
numerical statistical techniques on real time data measurements; for 
illustration
we have selected stratus clouds.

Restricting ourselves to cloud physics and fractal geometry ideas, 
leads to many questions, such as on the perimeter-area relationship of rain and cloud areas
\cite{lovejoy1}, the fractal dimension of their shape or ground projection
\cite{lovejoy2} or modelization of fractally homogeneous turbulence
\cite{theory}. The cloud inner structure, content, temperature, life time and
effects on ground level phenomena or features are of constant 
interest and prone
to physical modelisation \cite{nagel}. Recently, we reported about long-range
power-law correlations \cite{buda,kimaeeta} and multi-affine properties
\cite{kita} of stratus cloud liquid water fluctuations.

\subsection{Techniques of time series analysis}

The variety of systems that apparently display scaling properties ranges from
base-pair correlations in DNA and inter-beat intervals of the human heart, to
large, spatially extended geophysical processes, such as earthquakes, and signals
produced by complex systems, such as financial indices in economics. 
The current
paradigm is that these systems obey ``universal'' laws due to the underlying
nonlinear dynamics and are independent of the microscopic details. 
Therefore one
can consider in meteorology to obtain characteristic quantities using the same
modern statistical physics methods as done in all of the other cases. Whence we
will focus on several techniques to describe the scaling properties of
meteorological time series, like the Fourier power spectrum of the signal
\cite{MalamudTurcotte}, detrended fluctuation analysis (DFA) method 
\cite{DNADFA}
and its extension local DFA method \cite{buda}, and multi-affine and singularity
analysis \cite{kita,DM94}. One can go beyond these methods using wavelet
techniques \cite{wavelets}  or Zipf diagrams \cite{zipfbook,mazipf,mkzipf}. The
Fokker-Planck equation \cite{friedrich1} for describing the liquid water path
\cite{kijats}, which is studied  here below, is also of interest.

\section{Experimental techniques and data acquisition}

Quantitative observations of the atmosphere are made in many different ways.
Experimental/observational techniques to study the atmosphere rely on physical
principles. One important type of observational techniques is that of 
{\it remote
sensing}, which depends on the detection of electromagnetic radiation emitted,
scattered or transmitted by the atmosphere. The instruments can be placed at
aircrafts, on balloons or on the ground. Remote-sensing techniques can be
divided into {\it passive} and {\it active} types. In passive remote sensing,
the radiation measured is of natural origin, for example the thermal radiation
emitted by the atmosphere, or solar radiation transmitted or scattered by the
atmosphere. Most space-born remote sensing methods are passive. In 
active remote
sensing, a transmitter, e.g. a radar, is used to direct pulses of 
radiation into
the atmosphere, where they are scattered by atmospheric molecules, aerosols or
inhomogeneities in the atmospheric structure. Some of the scattered 
radiation is
then detected by some receiver. Each of these techniques has its advantages and
disadvantages. Remote sensing from satellites can give near-global 
coverage, but
can provide only averaged values of the measured quantity over large 
regions, of
order of hundreds of kilometers in horizontal extent and several kilometers in
the vertical direction. Satellite instruments are expensive to put 
into orbit and
cannot usually be repaired if they fail. Ground-based radars can provide data
with very high vertical resolution (by measuring small differences in the time
delays of the return pulses), but only above the radar site.

For illustrative purposes, we will use microwave radiometer data obtained from the
Department of Energy (DOE) Atmospheric Radiation Measurement (ARM) program
\cite{arm} site located at the Southern Great Plains (SGP) central facility
\cite{web}. For detailed presentation of other remote sensing techniques the
reader can consult Andrews \cite{andrews} and/or Rees \cite{rees}.

In this study we focus on stratus cloud data. For comparison the
cumulus cloud
scale is too small to be represented individually in today's numerical models
\cite{garrattbook}. Due to their relatively small sizes cumulus clouds produce
short time series when remote sensing measurements are applied. Therefore they
are not particularly suitable for the techniques that are outlined in this
report. However their role in the transport of heat, moisture and momentum must
be considered in numerical models.

The data used in this study are the vertical column amounts of cloud 
liquid water
that are retrieved from the radiances, recorded as brightness temperatures,
measured with a Radiometrics Model WVR-1100 microwave radiometer at frequencies
of 23.8 and 31.4 GHz \cite{westwater78,westwater,errorref}. The microwave
radiometer is equipped with a Gaussian-lensed microwave antenna whose 
small-angle
receiving cone is steered with a rotating flat mirror \cite{web}. The microwave
radiometer is located at the DOE ARM program SGP central facility and is
operated in
the vertically pointing mode. In this mode the radiometer makes sequential 1 s
radiance measurements in each of the two channels while pointing vertically
upward into the atmosphere. After collecting these radiances the radiometer
mirror is rotated to view a blackbody reference target. For each of the two
channels the radiometer records the radiance from the reference immediately
followed by a measurement of a combined radiance from the reference and a
calibrated noise diode. This measurement cycle is repeated once every 20 s.

A shorter measurement cycle does not necessarily lead to a larger number of
independent samples. For example, clouds at 2 km  altitude moving 
at 10 $\rm m
\,s^{-1}$ take 15 s to advect through a radiometer field-of-view of 
approximately
5$^\circ.$ Note that the 1 s sky radiance integration time ensures that the
retrieved quantities correspond to a specific column of cloud above the
instrument, as opposed to some longer time average of the cloud 
properties in the
column above the instrument. The field of view of the microwave radiometer is
5.7$^\circ$ at 23.8 GHz and 4.6$^\circ$ at 31.4 GHz.

Based on a standard model \cite{westwater78,errorref} (see Appendix), the
microwave radiometer measurements at the two frequency channels of 23.8 and 31.4 GHz
are used to obtain time series of  liquid water path (LWP) that 
corresponds to the total amount of liquid water within the vertical column
of the atmosphere that has been remotely sounded. The
error for the liquid water retrieval is estimated to be less than about
0.005~$g/cm^2$ \cite{errorref}.

The liquid water path (LWP) data $y(t)$ considered in this study are 
obtained on
April 3-5, 1998 and are shown in Fig. 1a.

\begin{figure} \centering 
\includegraphics[width=.48\textwidth]{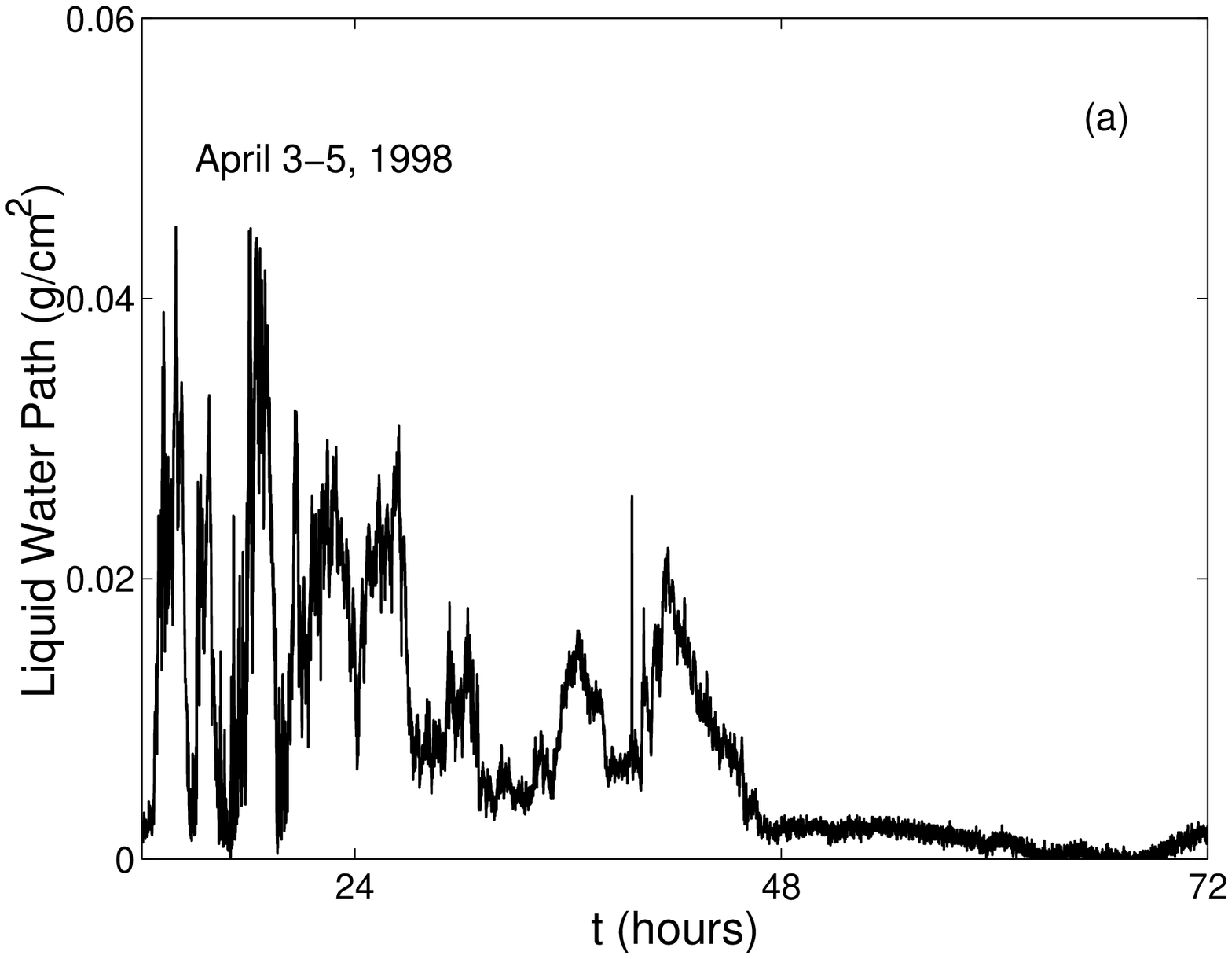} \hfill
\includegraphics[width=.48\textwidth]{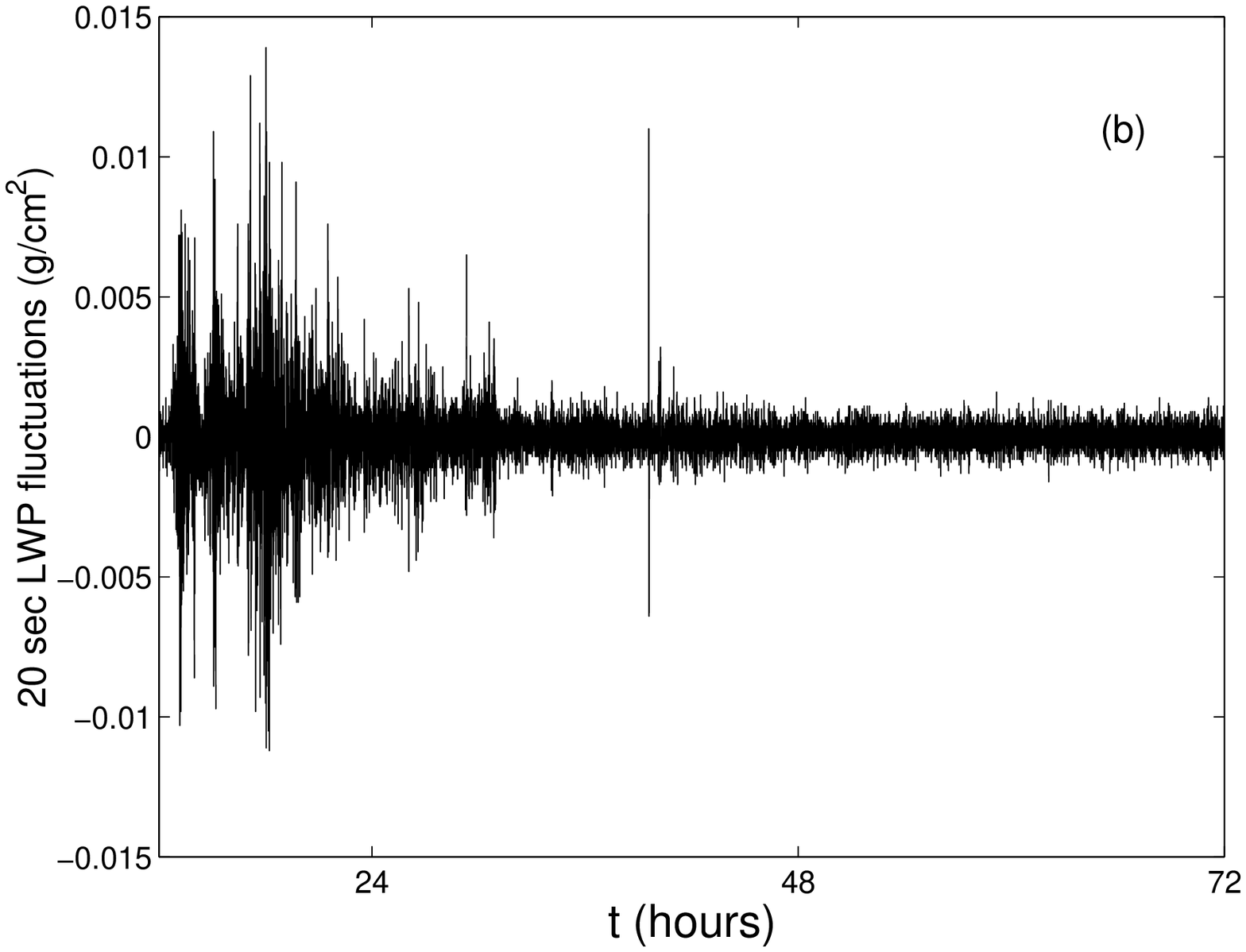} \vfill
\caption[]{{\bf (a)} Time dependence of liquid water path as obtained at the ARM
Southern Great Plains site with time resolution of 20~s during the period from
April 3 to 5, 1998. The time series contains $N=10740$ data points. On x-axis
t=24~h marks midnight on April 3, t=48~h corresponds to midnight on April 4 and
t=72~h corresponds to midnight on April 5, 1998. {\bf (b)} Small-scale gradient
field of the LWP signal, e.g. fluctuations of LWP for a time interval 
equal to the
discretization step of the measurements.} 
\label{eps1} 
\end{figure}

\section{Nonstationarity and Spectral density}

Fluctuations of the LWP signal $y(t)$ (data in Fig. 1a) are plotted in Fig. 1b
for the time interval equal to the discretization step of the data, 
i.e. $\Delta
t=20$~sec. This time series is also called the small-scale gradient 
field. Other
values of time intervals to study fluctuations of a signal can be of 
interest to
search for changes in the type and strength of the correlations
\cite{friedrich2}. This approach will not be pursued here.

One approach to test the type of the LWP fluctuations is to estimate the
nonstationarity of the signal. The power spectral density $S(f)$ of the time
series $y(t)$ is defined as the Fourier transform of the signal. For supposedly
self-affine signals $S(f)$ is expected to follow a power-law 
dependence in terms
of the frequency $f$,

\begin{equation} S(f) ~\sim ~ f^{-\beta}. \end{equation}

Equation (1) allows one to put the phenomena that produce the time series into
the class of {\it self-affine} phenomena.

It has been argued \cite{mandel,extreme} that the spectral exponent $\beta$
contains information about the degree of stationarity of the signal $y(t)$.
Depending on the value of $\beta$ the time series is called stationary or not;
for $\beta<1$, the signal is statistically invariant by transition in 
time, thus
called stationary. If $\beta>1$, the signal is nonstationary. In addition, if
$\beta<3$ the increments of the signal form a stationary series, in particular
the small-scale gradient field is stationary. Many
geophysical fields are nonstationary with stationary increments ($1<\beta<3$)
over some scaling range. The upper bound of the nonstationary regime 
is required
to keep the field values within their physically accessible range by 
limiting the
amplitude of the large scale fluctuations, which corresponds to a 
flatter part of
the spectrum at low frequencies.

Brownian motion is characterized by $\beta$ = 2, and  white noise 
by $\beta$ =
0. Indeed the Brownian motion or random walk $z(x)$ is a classical example of a
nonstationary process. We know that its variance $<z^2(x)>$ is proportional to
$x$, which proves the nonstationarity in the one-point statistics. 
However, in the
framework of two-point statistics, this result has a different 
interpretation. The
variance of the ``increment'' $z(x+\xi)-z(x)$ increases linearly with $\xi$,
independently of $x$, which is an indication of the stationarity of the
increments.

The range over which the $\beta$ exponent is well defined in Eq. (1) indicates
the range over which the scaling properties of the time series are 
invariant. The
power spectral density $S(f)$ of the liquid water path data measured on April
3-4, 1998 is shown in Fig. 2. The spectral exponent $\beta=1.56\pm 0.03$
indicates a nonstationary time series.

\begin{figure}
\includegraphics[width=.9\textwidth]{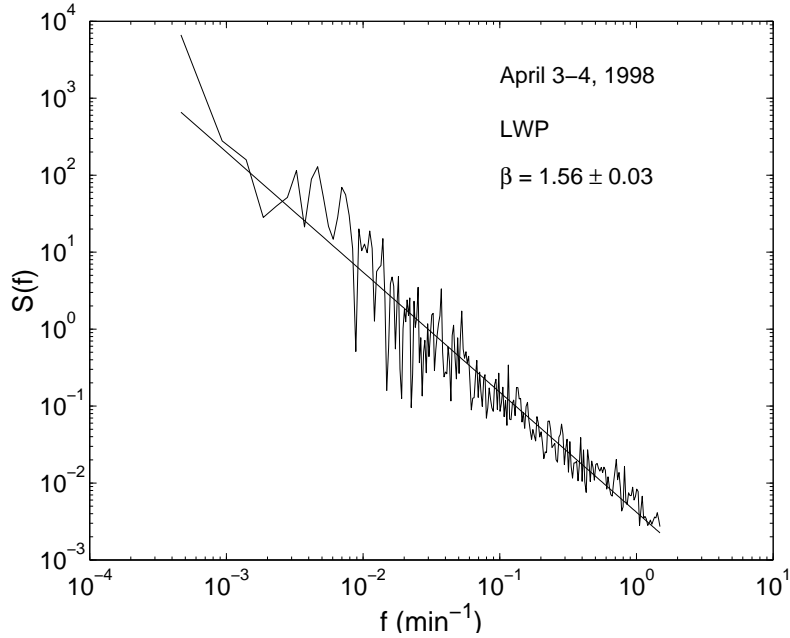}
\caption[]{Power spectral density for data measured on April 3-4, 1998.}
\label{eps2} 
\end{figure}

\section{Roughness and Detrended Fluctuation Analysis}

The fractal dimension \cite{schroeder,addison,falconer,west} $D$ is 
often used to
characterize the roughness of profiles \cite{roughness}. Several 
methods are used
for measuring $D$, like the box counting method, though not quite 
efficient; many
others are found in the literature as seen in
\cite{schroeder,addison,falconer,west} and here below. For topologically one
dimensional systems, the fractal dimension $D$ is related to the 
exponent $\beta$
by

\begin{equation} \beta = 5-2D. \end{equation}

Another ''measure'' of the signal roughness is sometimes given by the 
Hurst $Hu$
exponent, first defined in the ''rescale range theory'' (of Hurst 
\cite{Hu4,Hu5})
who suggested a method to estimate the persistence of the Nile floods and
droughts. The Hurst method consists of listing the differences between the
observed value at a discrete time $t$ over an interval with size $N$ 
on which the
mean has been taken. The upper ($y_M$) and lower ($y_m$) values in 
that interval
define the range $R_N = y_M - y_m$. The root mean square deviation $S_N$ being
also calculated, the ''rescaled range'' is $R_N/S_N$ is expected to behave like
$N^{Hu}$. This means that for a (discrete) self-affine signal $y(t)$, the
neighborhood of a particular point on the signal can be rescaled by a 
factor $b$
using the roughness (or Hurst \cite{addison,falconer}) exponent $Hu$ 
and defining
the new signal $ b^{-Hu} y(bt)$. For the exponent value $Hu$, the frequency
dependence of the signal so obtained should be undistinguishable from the
original one, i.e. $y(t)$.

The roughness (Hurst) exponent $Hu$ can be calculated from the height-height
correlation function $c_1(\tau)$ or first order structure function 
that supposed
to behave like \begin{equation} c_1(\tau) = {\langle {| y(t_{i+r})-y(t_i) |}
\rangle}_{\tau} \sim {\tau}^{H_1} \end{equation} whereas 
\begin{equation} Hu = 1
+ H_1, \end{equation} rather than from the box counting method. For a {\it
persistent} signal, $H_1>1/2$; for an {\it anti-persistent} signal, $H_1<1/2$.
Flandrin has theoretically proved \cite{Flandrin} that

\begin{equation} \beta = 2 Hu - 1, \end{equation} thus $\beta$ = 1+2 
$H_1$. This
implies that the classical random walk (Brownian motion) is such that $Hu=3/2$.
It is clear that

\begin{equation} D = 3 - Hu. \end{equation}

Fractional Brownian motion values in other fields \cite{3a,3b,nvma} are
practically found to lie between $1$ and $2$. Since a white noise is a truly
random process, it can be concluded that $Hu=1.5$ implies an uncorrelated time
series \cite{west}.

Thus $D>1.5$, or $Hu<1.5$ implies antipersistence and $D<1.5$, or $Hu>1.5$
implies persistence. From preimposed $Hu$ values of a fractional 
Brownian motion
series, it is found that the equality here above usually holds true in a very
limited range and $\beta$ only slowly converges toward the value $Hu$
\cite{MalamudTurcotte,Turcottebook}.

The above exponents and parameters can be obtained within the detrended
fluctuations analysis (DFA) method \cite{DNADFA}. The DFA method is a tool used
for sorting out correlations in a self-affine time series with stationary
increments \cite{nvma,ijmpc,hu}. It provides a simple quantitative parameter -
the scaling exponent $\alpha$, which is a signature of the 
correlation properties
of the signal. The advantages of DFA over many methods are that it permits
detection of long-range correlations embedded in seemingly non-stationary time
series, and also that inherent trends are avoided at all time scales. The DFA
technique consists in dividing a time series $y(t)$ of length $N$ into $N/\tau$
nonoverlapping boxes (called also windows), each containing $\tau$ points
\cite{DNADFA}. The local trend $z(n)$ in each box is defined to be the ordinate
of a linear least-square fit of the data points in that box. The detrended
fluctuation function $F^2(\tau)$ is then calculated following:

\begin{equation} F^2(\tau) = {1 \over \tau } {\sum_{n=k\tau+1}^{(k+1)\tau}
{\left[y(n)- z(n)\right]}^2} \qquad k=0,1,2,\dots,\left(\frac{N}{\tau}-1\right)
\end{equation}

Averaging $F^2(\tau)$ over the $N/\tau$ intervals gives the mean-square
fluctuations

\begin{equation} <F^2(\tau)>^{1/2} \sim \tau^{\alpha}. \end{equation}

The DFA exponent $\alpha$ is obtained from the power law scaling of 
the function
$<F^2(\tau)>^{1/2}$ with $\tau$, and represents the correlation 
properties of the
signal: $\alpha=1/2$ indicates that the changes in the values of a time series
are random and, therefore, uncorrelated with each other. If $\alpha<1/2$ the
signal is anti-persistent (anti-correlated), while $\alpha>1/2$ indicate 
positive persistency (correlation) in the signal.

Results of the DFA analysis of liquid water path data measured on 
April 3-4, 1998
are plotted in Fig. 3a. The DFA function is close to a power law with 
an exponent
$\alpha = 0.34 \pm 0.01$ holding from 3 to 60 minutes. This scaling range is
somewhat shorter than the 150~min scaling range we obtained 
\cite{kimaeeta} for a
stratus cloud during the period Jan. 9-14, 1998 at the ARM SGP site. 
A crossover
to $\alpha=0.50 \pm 0.01$ is readily seen for longer correlation 
times \cite{hu}
to about 2 h, after which the statistics of the DFA function is not 
reliable. One
should note that for cloud data the lower limit of the scaling range is
determined by the resolution and discretization steps of the 
measurements. Since
such clouds move at an average speed of {\it ca.} 10 m/s and the instrument is
always directed toward the same point of the atmosphere, the $20 s$
discretization step is chosen to ensure ergodic sampling for an about 
$5^{\circ}$
observation angle of the instrument. The upper scaling range limit 
depends on the
cloud life time.

The value of $\alpha \approx 0.3$ can be interpreted as the $H_1$ parameter of
the multifractal analysis of liquid water content \cite{DM94} and of 
liquid water
path \cite{kita}.  The existence of a
crossover suggests two types of correlated events as in classical fracture
processes: (i) On one hand, the nucleation part and the growth of diluted
droplets occur in ''more gas-like regions''. This process is typically slow and
is governed by long range Brownian-like fluctuations; it is expected 
to follow an
Eden model-like \cite{eden} growth, with a trivial scaling exponent, as $\alpha
=$ 0.5 (Fig. 3b); (ii) The faster processes with more Levy-like 
fluctuations are
those which link together various fracturing parts of the cloud, and are
necessarily antipersistent as long as the cloud remains 
thermodynamically stable;
they occur at shorter correlation times, and govern the cloud breaking final
regime as in any percolation process \cite{StanleyPTbook}, - with an intrinsic
non-trivial scaling exponent $\sim$ 0.3.

\begin{figure} \centering 
\includegraphics[width=.48\textwidth]{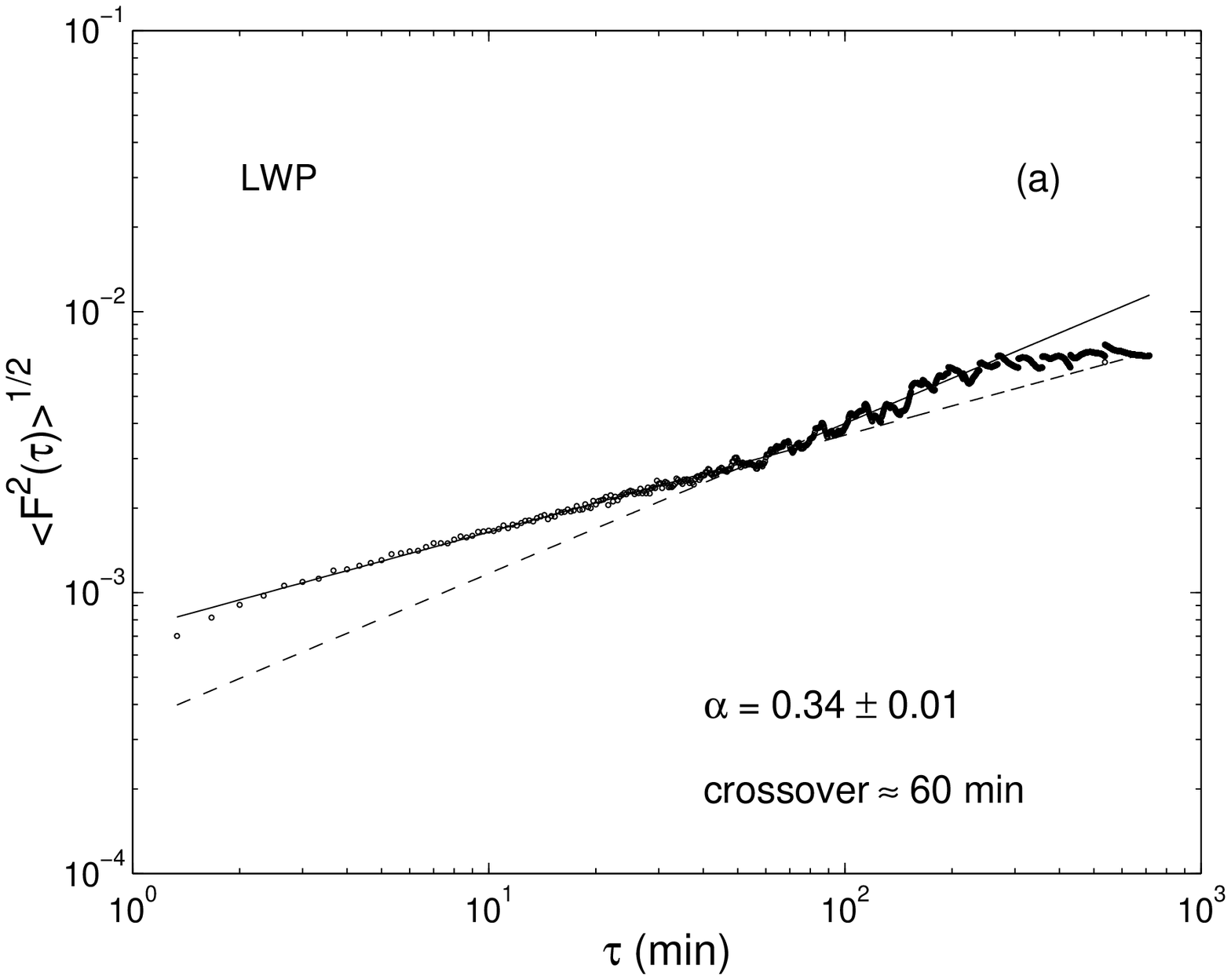} \hfill
\includegraphics[width=.48\textwidth]{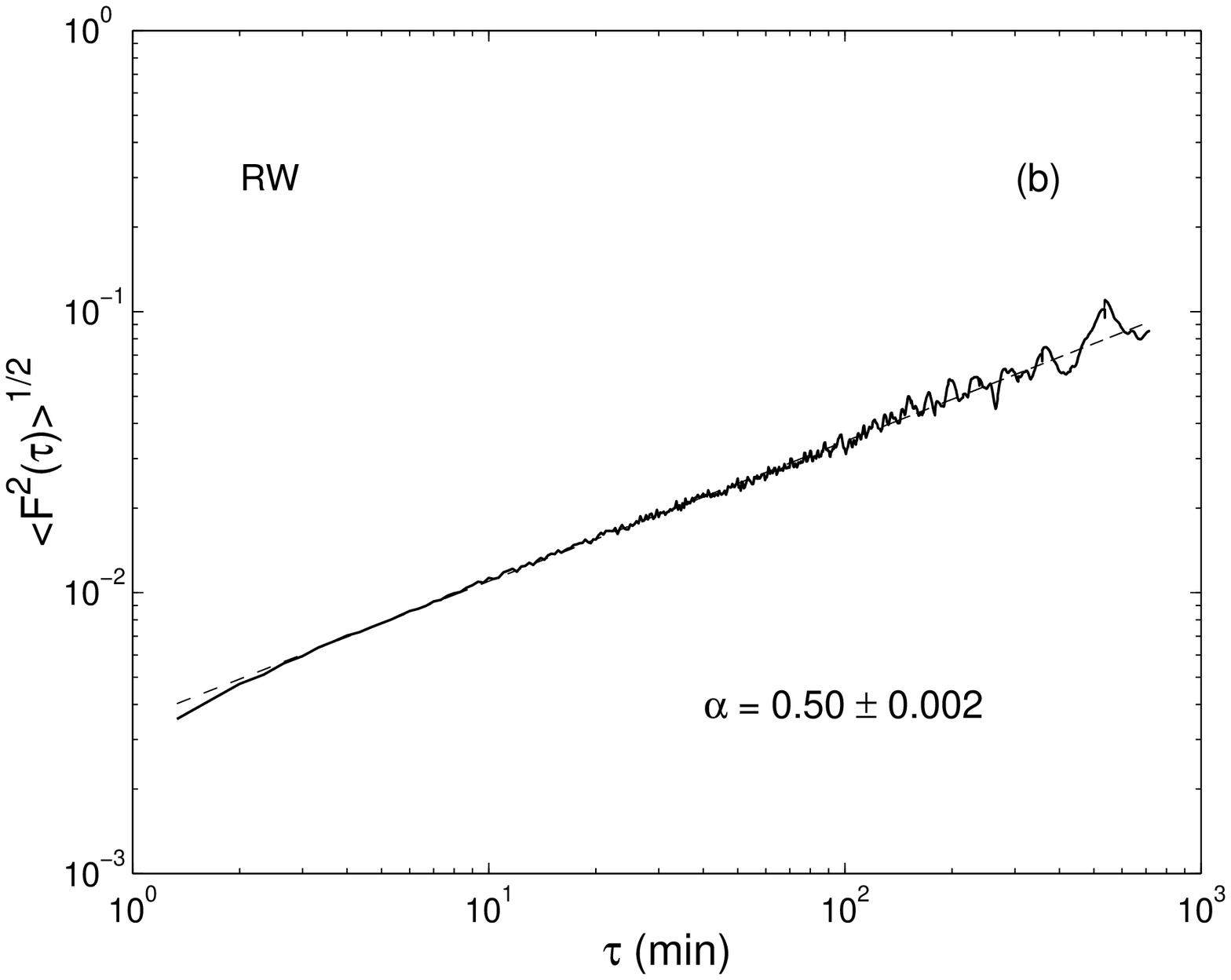} \vfill
\caption[]{(a) Detrended fluctuation function $<F^2(\tau)>^{1/2}$ for 
data measured
on April 3-4, 1998. (b) DFA-function for Brownian walk signal scales with
$\alpha=0.50\pm0.01$ and is plotted for comparison.} 
\label{eps3}
\end{figure}

Several remarks are in order. Recently a rigorous relation 
between detrended fluctuation analysis and 
power spectral density analysis for stochastic processes is established \cite{heneghan}.
Thus, if the two scaling exponents $\alpha$ and $\beta$
are well defined,  $\beta = 2 \alpha +1$ holds 
for $0<\alpha<1$ ($1<\beta<3$) for fractional Brownian walks \cite{Turcottebook,Monin}. 
establishing the relation between detrended fluctuation analysis and 
power spectral density analysis for stochastic processes

In terms of
the exponents ($\alpha$ and $\beta$) of the signal, we can talk about 
pink noise
$\alpha=0$ ($\beta=1$), brown noise $\alpha=1/2$ ($\beta=2$) or black noise
$\alpha>1/2$ ($\beta>2$) \cite{schroeder}. Black noise is related to 
persistence.
In contrast, inertial subrange turbulence for which $\beta=5/3$ gives
$\alpha=1/3$ \cite{turbulence}, which places it in the antipersistence regime.

The two scaling exponents $\alpha$ and $\beta$ for the liquid water path signal
are only approximately close to fulfilling the relation $\beta=2\alpha+1$. This
can be interpreted to be due to the peculiarities of the spectral method
\cite{panter}. In general, the Fourier transform is inadequate for 
non-stationary
signals. Also it is sensitive to possible trends in the data. There 
are different
techniques suggested to correct these deficiencies of the spectral method
\cite{priestly,percival}, like detrending the data before taking the Fourier
transform. However, this may lead to questions about the accuracy of 
the spectral
exponent \cite{pelletier}.

\section{Time dependence of the correlations}

In previous section we study the type of correlations that exist in the liquid
water path signal measured during cloudy atmospheric conditions, on April 3-4,
1998. Here we focus on the evolution of these correlations during the same time
interval but also continuing on the next day, April 5, when the stratus cloud
disappears. In doing so we can further study the influence of the time lag on
correlations in the signal.

In order to probe the existence of so called {\it locally correlated} and {\it
decorrelated} sequences \cite{nvma}, one can construct a so-called observation
box with a certain width, $\tau$, place the box at the beginning of the data,
calculate $\alpha$ for the data in that box, move the box by 
$\Delta\tau$ toward
the right along the signal sequence, calculate $\alpha$ in that box, 
a.s.o. up to
the $N$-th point of the available data. A time dependent $\alpha$ 
exponent may be
expected.

We apply this technique to the liquid water path data signal and the result is
shown in Fig. 4. For this illustration we have chosen two window 
sizes, i.e. 4~h
and 6~h, moving the window with a step of $\Delta\tau=1$~h. Since the value of
{\it local} $\alpha$ can only be known after all data points are taken into
account in a box, the reported value corresponds to that at the upper most time
value for that given box in Fig. 4. One clearly observes that the $\alpha$
exponent value does not vary much when the value of $\tau$ and $\Delta\tau$ are
changed. As could be expected there is more roughness if the box is 
narrower. The
local $\alpha$ exponent value is always significantly below 1/2. By 
analogy with
financial and biological studies, this is interpreted as a phenomenon 
related to
the $fractional$ $Brownian$ $motion$ process mentioned above.

\begin{figure}
\includegraphics[width=.9\textwidth]{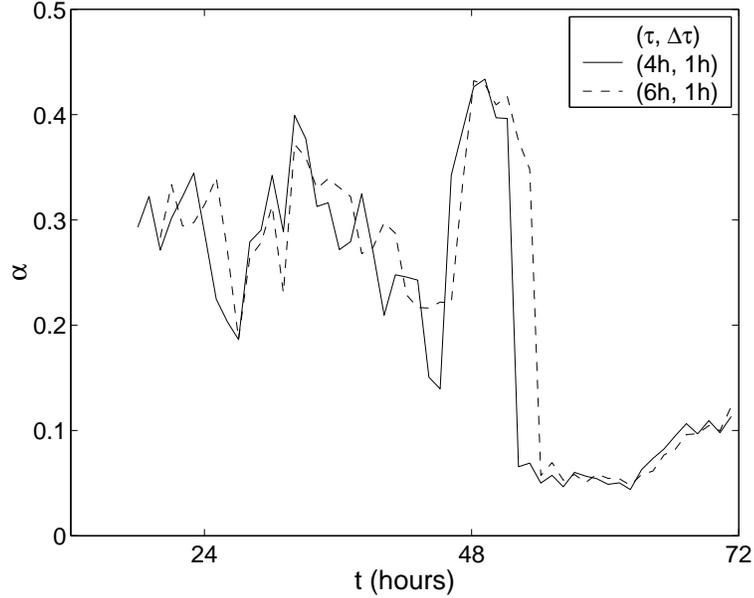}
\caption[]{Local $alpha$-exponent from DFA analysis for data in Fig. 1a.}
\label{eps4} \end{figure}

The results from this local DFA analysis applied to LWP data (Fig. 4) indicate
two well defined regions of scaling with different values of 
$\alpha$. The first
region corresponds to the first two days when a thick stratus cloud 
existed. The
average value of the local scaling exponent over this period is $\alpha = 0.34
\pm 0.01$; it is followed by a sharp rise to 0.5, then by a sharp drop below
$\alpha = 0.1$ when there is a clear sky day. These values of local 
$\alpha$ are
well defined for a scaling time (range) interval extending between 2 and 25
minutes for the various $\tau$ and $\Delta\tau$ combinations. The value of
$\alpha$, close to $0.3$, indicates a very large antipersistence, thus a set of
fluctuations tending to induce a great stability of the system and great
antipersistence of the prevailing meteorology, - in contrast to the 
case in which
there would be a persistence of the system which would be dragged out of
equilibrium; it would equally imply good predictability. This implies that
specific fluctuation correlation dynamics could be usefully inserted as
ingredients in {\it ad hoc} models.

The appearance of a patch of clouds and clear sky following a period of thick
stratus can be interpreted as a non equilibrium transition. The $\alpha = 1/2$
value in financial fluctuations \cite{nvma} was observed to indicate 
a period of
relative economic calm. The appropriately called thunderstorm of activities and
other bubble explosions in the financial field correspond to a value different
from $1/2$ \cite{ndub}. Thus we emphasize here that stable states can occur for
$\alpha$ values that do not correspond to the Brownian $1/2$ value. 
We conclude
that the fluctuation behavior is an observational feature more 
important than the
peak appearance in the raw data. Moreover, from a fundamental point of view, it
seems that the variations of $\alpha$ are as important as the value itself
\cite{nvma}. From the point of view of predictability, $\alpha$ values
significantly different from $1/2$ are to be preferred because such 
values imply
a great degree of predictability and stability of the system.

\section{Multi-affinity and Intermittency}

The variations in the local $\alpha$-exponent suggest that the nature of the
correlations changes with time. As a consequence the evolution of the 
time series
can be decomposed into successive persistent and anti-persistent sequences
\cite{nvma}, and multi-affine behavior can be expected. Multi-affine properties
of a time dependent signal $y(t)$ are described by the so-called ``q-th'' order
structure functions

\begin{equation} c_q = \left<|y(t_{i+r}) - y(t_i)|^q \right> \qquad 
i=1,2, \dots
, N -r \end{equation} where the average is taken over all possible pairs of points
that are apart from each other a distance $\tau=y(t_{i+r})-y(t_i)$.

Assuming a power law dependence of the structure function, the $H(q)$ 
spectrum is
defined through the relation \cite{DMWC96,MDWC97}

\begin{equation} c_q(\tau)\sim \tau^{qH(q)} \qquad q\ge 0 
\label{hq}
\end{equation}

The {\it intermittency} of the signal can be studied through the so-called
singular measure analysis. The first step that this technique require 
is defining
a basic measure $\varepsilon(1;l)$ as

\begin{equation} \varepsilon(1;l)=\frac{|\Delta y(1;l)|}{<\Delta 
y(1;l)>}, \qquad
l=0,1, \dots, N -1 \end{equation} where $\Delta 
y(1;l)=y(t_{i+1})-y(t_i)$ is the
small-scale gradient field and

\begin{equation}<\Delta y(1;l)>= \frac{1}{N}\sum_{l=0}^{N-1}|\Delta y(1;l)|.
\label{ave} \end{equation}

This is indeed deriving a stationary nonnegative field from a 
nonstationary data
and this is the simplest procedure to do so. Other techniques involve
''fractional'' derivatives \cite{schmitt} or second derivatives \cite{tessier}.
Also one can consider taking squares \cite{mene} rather than the 
absolute values
but that leads to a linear relation between the exponents of these 
two measures.
It is argued elsewhere \cite{lavallee} that the details of the procedure do
not influence the final results of the singularity analysis.

We use a spatial/temporal average in Eq. (\ref{ave})
rather than an ensemble average, thus making an ergodicity assumption
\cite{ergod,ergodmf} that is our only recourse in empirical data analysis.

Next we define a series of ever more coarse-grained and ever shorter fields
$\varepsilon(r;l)$ where $0<l<N-r$ and $r=1,2,4,\dots,N=2^m$. Thus the average
measure in the interval $[l;l+r]$ is

\begin{equation}\varepsilon(r;l)=\frac{1}{r}\sum_{l'=l}^{l+r-1} 
\varepsilon(1;l')
\qquad l=0, \dots , N - r \end{equation}

The scaling properties of the generating function are then searched for through
the equation

\begin{equation} \chi_q(\tau)=<\varepsilon(r;l)^q>\sim \tau^{-K(q)} ,\quad q\ge
0, 
\label{kq}
\end{equation} with $\tau=y(t_{i+r})-y(t_i)$.

It should be noted that the intermittency of a signal is related to 
existence of
extreme events, thus a distribution of events away from a Gaussian 
distribution,
in the evolution of the process that has generated the data. If the 
tails of the
distribution function follow a power law, then the scaling exponent defines the
critical order value after which the statistical moments of the signal diverge
\cite{extreme}. Therefore it is of interest to probe the distribution of the
fluctuations of a time dependent signal $y(t)$ prior investigating its
intermittency. The distribution of the fluctuations of liquid water path signal
measured on April 3-4, 1998 at the ARM Southern Great Plains site is shown in
Fig. 5.

\begin{figure}
\includegraphics[width=.9\textwidth]{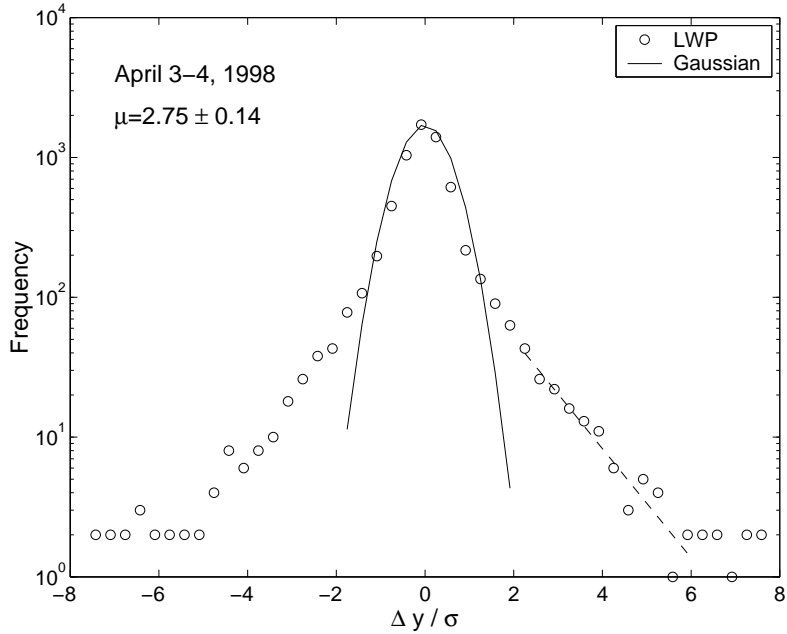}
\caption[]{Distribution of the frequency of LWP fluctuations $\Delta
y/\sigma=(y(t_{i+1})-y(t_i))/\sigma$, where $\sigma=0.0011 g/cm^2$ is the
standard deviation of the fluctuations for LWP signal measured on 
April 3-4, 1998
(data in Fig. 1a)} 
 \end{figure}

The frequency distribution is not Gaussian but is rather symmetrical. The tails
of the distribution follow a power law

\begin{equation} P(x) \sim \frac{1}{x^\mu} \end{equation} with an exponent
$\mu=2.75\pm0.12$ away from the Gaussian $\mu=2$ value. This scaling law gives
support to the argument in favor of the existence of self-affine properties, as
established in section 4 for the LWP signal, when applying the DFA method. The
extreme events that form the tails of the probability distribution also
characterize the intermittency of the signal. In Fig. 6 the multi-fractal
properties of the LWP signal are expressed by two sets of scaling functions,
the $H(q)$ hierarchy of functions describing the roughness of the signal and the $K(q)$
hierarchy of functions describing its intermittency as defined in 
Eq.(\ref{hq}) and Eq.
(\ref{kq}) respectively.  For $q=1$, $H(1)$ is the value
that is given by the DFA analysis.

\begin{figure}
\includegraphics[width=.9\textwidth]{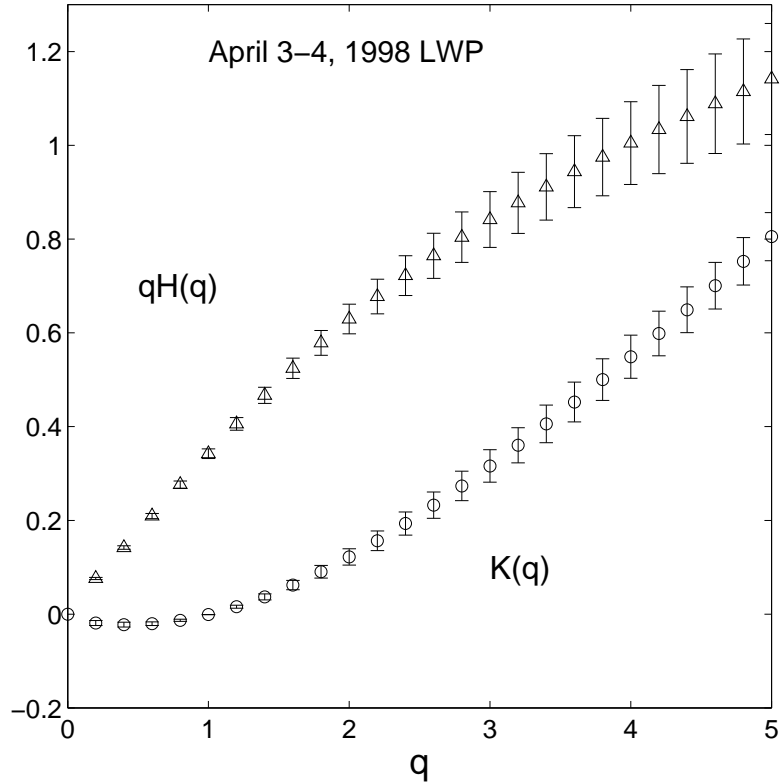}
\caption[]{The $H(q)$ and $K(q)$ functions for the LWP data obtained on 
April 3-4,
1998. } \end{figure}

\section{Conclusions}

Scaling properties of the liquid water path in stratus clouds have 
been analyzed
to demonstrate the application of several methods of statistical physics for
analyzing data in atmospheric sciences, and more generally in geophysics. We have
found that the breaking up of a stratus cloud is related to changes in the type of
correlations in the fluctuations of the signal, that represents the total
vertical amount of liquid water in the stratus cloud. We have demonstrated that
the correlations of LWP fluctuations exist indeed in a more complex way than
usually known through their multi-affine dependence.

\section{Acknowledgements}

Thanks to Luc T. Wille for inviting us to present the above results 
and enticing
us into writing this report. Thanks to him and the State of Florida for some
financial support during the conference. This research was partially 
supported by Battelle grant number 327421-A-N4. We acknowledge 
collaboration of the U.S. Department of Energy as part of the Atmospheric 
Radiation Measurement Program.

\section{Appendix}

For
nonprecipitating
clouds, i.e., clouds having drops sufficiently small that scattering is
negligible, measurements of the microwave radiometer brightness
temperature $T_{{\rm B}\omega}$ can be mapped onto an opacity 
$\nu_{\omega}$ parameter by

\begin{equation}
\nu_{\omega} = \ln\left[{ {(T_{\rm mr} - T_{\rm c})} \over {(T_{\rm mr} -
T_{{\rm B}\omega})} }\right],
\label{ret1}
\end{equation}
where $T_{\rm c}$ is the cosmic background ``big bang'' brightness
temperature equal to 2.8~K and $T_{\rm mr}$ is an estimated ``mean radiating
temperature'' of the atmosphere.

Writing $\nu_{\omega}$ in terms of atmospheric constituents, we have

\begin{equation}
\nu_{\omega} = \kappa_{{\rm V}\omega}V + \kappa_{{\rm L}\omega}L + 
\nu_{{\rm d}\omega},
\label{ret2}
\end{equation}
where $\kappa_{{\rm V}\omega}$ and $\kappa_{{\rm L}\omega}$ are {\it 
water vapor and liquid water path}-averaged mass
absorption coefficients and $\nu_{{\rm d}\omega}$ is the
absorption by dry atmosphere constituents (e.g., oxygen). Next, define

\begin{equation}
\nu_{\omega}^\ast = \nu_{\omega} - \nu_{{\rm d}\omega} = \ln\left[{ 
{(T_{\rm mr}
- T_{\rm c})}
\over {(T_{\rm mr} - T_{{\rm B}\omega})} }\right] - \nu_{{\rm d}\omega}.
\end{equation}
The 23.8~GHz channel is sensitive primarily to water
vapor while
the 31.4~GHz channel is  sensitive
primarily to cloud liquid  water. Therefore  two equations for the
opacity can be written for each
  frequency and then solved for the two unknowns $L$ and $V$, i.e.

\begin{equation}
L = l_1\nu^\ast_{\omega_1} + l_2\nu^\ast_{\omega_2} \qquad \qquad (LWP)
\end{equation}

and
\begin{equation}
V = v_1\nu^\ast_{\omega_1} + v_2\nu^\ast_{\omega_2}, \qquad \qquad (WVP)
\end{equation}

where

\begin{equation}
l_1 = - \left(\kappa_{{\rm L}\omega_2}{{\kappa_{{\rm V}\omega_1}} \over
{\kappa_{{\rm V}\omega_2}}} - \kappa_{{\rm L}\omega_1}\right)^{-1},
\end{equation}

\begin{equation}
l_2 = \left(\kappa_{{\rm L}\omega_2} -
\kappa_{{\rm L}\omega_1}{{\kappa_{{\rm V}\omega_2}}\over {\kappa_{{\rm
V}\omega_1}}}\right)^{-1},
\end{equation}

\begin{equation}
v_1 = \left(\kappa_{{\rm V}\omega_1} -
\kappa_{{\rm V}\omega_2}{{\kappa_{{\rm L}\omega_1}}\over {\kappa_{{\rm
L}\omega_2}}}\right)^{-1},
\end{equation}

\begin{equation}
v_2 = - \left(\kappa_{{\rm V}\omega_1}{{\kappa_{{\rm L}\omega_2}}\over
{\kappa_{{\rm L}\omega_1}}} -
\kappa_{{\rm V}\omega_2}\right)^{-1}.
\end{equation}

\end{document}